\documentclass[aps,prc,twocolumn,showpacs,superscriptaddress,letter]{revtex4-1}  
\usepackage{graphicx}  
\usepackage{bm}        
\usepackage{amssymb}   
\usepackage{amsmath}
\usepackage{longtable}
\usepackage{dcolumn}
\usepackage{multirow}
\usepackage{verbatim}

\begin{document}

\title{High-resolution (p,t) reaction measurements as spectroscopic tests of {\it ab-initio} theory in the mid $pf$-shell}

\author{K.G.~Leach}\email{kleach@mines.edu}\address{Department of Physics, Colorado School of Mines, Golden, Colorado 80401, USA}\address{Department of Physics, University of Guelph, Guelph, Ontario N1G 2W1, Canada}\address{TRIUMF, 4004 Wesbrook Mall, Vancouver, British Columbia V6T 2A3, Canada}
\author{J.D.~Holt}\affiliation{TRIUMF, 4004 Wesbrook Mall, Vancouver, British Columbia V6T 2A3, Canada}
\author{P.E.~Garrett}\affiliation{Department of Physics, University of Guelph, Guelph, Ontario N1G 2W1, Canada}\affiliation{Excellence Cluster Universe, Boltzmannstra\ss{}e 2, D-85748 Garching, Germany}
\author{S.R.~Stroberg}\altaffiliation{Present address: Department of Physics and Astronomy, University of Washington, Seattle, WA 98195, USA}\affiliation{TRIUMF, 4004 Wesbrook Mall, Vancouver, British Columbia V6T 2A3, Canada}
\author{G.C.~Ball}\affiliation{TRIUMF, 4004 Wesbrook Mall, Vancouver, British Columbia V6T 2A3, Canada}
\author{P.C.~Bender}\altaffiliation{Present address: Department of Physics, University of Massachusetts at Lowell, Lowell, Massachusetts 01854, USA}\affiliation{TRIUMF, 4004 Wesbrook Mall, Vancouver, British Columbia V6T 2A3, Canada}
\author{V.~Bildstein}\affiliation{Department of Physics, University of Guelph, Guelph, Ontario N1G 2W1, Canada}
\author{A.~Diaz~Varela}\affiliation{Department of Physics, University of Guelph, Guelph, Ontario N1G 2W1, Canada}
\author{R.~Dunlop}\affiliation{Department of Physics, University of Guelph, Guelph, Ontario N1G 2W1, Canada}
\author{T.~Faestermann}\affiliation{Physik Department, Technische Universit\"at M\"unchen, D-85748 Garching, Germany}
\author{B.~Hadinia}\affiliation{Department of Physics, University of Guelph, Guelph, Ontario N1G 2W1, Canada}
\author{R.~Hertenberger}\affiliation{Fakult\"at f\"ur Physik, Ludwig-Maximilians-Universit\"at M\"unchen, D-85748 Garching, Germany}
\author{D.S.~Jamieson}\altaffiliation{Present address: Department of Physics and Astronomy,Stony Brook University, Stony Brook, NY 11794, USA}\affiliation{Department of Physics, University of Guelph, Guelph, Ontario N1G 2W1, Canada}
\author{B.~Jigmeddorj}\affiliation{Department of Physics, University of Guelph, Guelph, Ontario N1G 2W1, Canada}
\author{R.~Kr\"ucken}\affiliation{TRIUMF, 4004 Wesbrook Mall, Vancouver, British Columbia V6T 2A3, Canada}\affiliation{Physik Department, Technische Universit\"at M\"unchen, D-85748 Garching, Germany}
\author{A.T.~Laffoley}\affiliation{Department of Physics, University of Guelph, Guelph, Ontario N1G 2W1, Canada}
\author{A.J.~Radich}\affiliation{Department of Physics, University of Guelph, Guelph, Ontario N1G 2W1, Canada}
\author{E.T.~Rand}\altaffiliation{Present Address: AECL Chalk River Laboratories, 286 Plant Rd. Stn 508A, Chalk River, Ontario K0J 1J0, Canada}\affiliation{Department of Physics, University of Guelph, Guelph, Ontario N1G 2W1, Canada}
\author{C.E.~Svensson}\affiliation{Department of Physics, University of Guelph, Guelph, Ontario N1G 2W1, Canada}
\author{S.~Triambak}\altaffiliation{Present address: University of the Western Cape, Robert Sobukwe Road, Bellville, 7535, Republic of South Africa}\affiliation{Department of Physics, University of Guelph, Guelph, Ontario N1G 2W1, Canada}\affiliation{TRIUMF, 4004 Wesbrook Mall, Vancouver, British Columbia V6T 2A3, Canada}
\author{H.-F.~Wirth}\affiliation{Fakult\"at f\"ur Physik, Ludwig-Maximilians-Universit\"at M\"unchen, D-85748 Garching, Germany}

\date{\today}

\begin{abstract}
Detailed spectroscopic measurements of excited states in $^{50}$Cr and $^{62}$Zn were performed using 24~MeV (p,t) transfer reactions on $^{52}$Cr and $^{64}$Zn, respectively.  In total, forty-five states in $^{50}$Cr and sixty-seven states in $^{62}$Zn were observed up to excitation energies of 5.5~MeV, including several previously unobserved states.  These experimental results are compared to {\it ab-initio} shell-model calculations using chiral effective field theory ($\chi$-EFT) with the valence-space in-medium similarity renormalization group (VS-IMSRG) method.  This comparison demonstrates good agreement in the level orderings with these new theoretical methods, albeit with a slight over binding in the calculations.  This work is part of a continued push to benchmark {\it ab-initio} theoretical techniques to nuclear structure data in $0^+\rightarrow0^+$ superallowed Fermi $\beta$ decay systems.

\end{abstract}

\pacs{21.60.Cs, 23.40.Bw, 24.10.Eq, 24.50.+g, 27.50.+e, 29.30.Ep}
\maketitle
\section{Introduction}

The development of a first-principles, or {\it ab-initio}, theoretical description of atomic nuclei is a central challenge for nuclear physics. This task is complicated by the combined difficulties of deriving nuclear interactions systematically and the complexity in solving the nuclear many-body problem. Developments in chiral effective field theory ($\chi$-EFT)~\cite{Mach11PR,Epel09RMP}, similarity renormalization group (SRG)~\cite{Bogn10PPNP}, and {\it ab-initio}  many-body techniques~\cite{Hage14RPP,Herg16PR} provide a unified picture for these efforts, where three-nucleon (3N)  forces  have  emerged  as  an  essential ingredient~\cite{Hebe15ARNPS}.

Particularly exciting is the prospect of using modern {\it ab-initio} theoretical techniques to provide a controlled framework for nuclear-structure corrections important for fundamental tests of the Standard Model with nuclei.  The primary advantage over standard phenomenological methods is the more transparent and systematic approach to the uncertainty estimates on the theoretical values~\cite{Car16}. Current applications include the calculation of $0\nu\beta\beta$ decay nuclear matrix elements~\cite{Enge16RPP}, dark matter scattering for direct detection experiments~\cite{Mene12WIMP}, and in the near future, isospin symmetry breaking (ISB) corrections for 0$^+\rightarrow$~0$^+$ superallowed Fermi $\beta$ decay~\cite{LeaCIPANP}.

However, many of the nuclei relevant for fundamental symmetry tests exist in medium-to-heavy mass, open-shell, and potentially deformed regions, which all present a particular challenge for {\it ab-initio} theory. Therefore confronting such methods with nuclear structure properties of relevant systems is a crucial first step towards reliable predictions.  This article presents the first attempt to study detailed spectroscopy of intermediate-mass mid-shell isotopes from an {\it ab-initio} approach in the superallowed $0^+\rightarrow0^+$ decay daughter nuclei $^{50}$Cr and $^{62}$Zn.

\begin{figure*}[t!]
\centering
\rotatebox{-90}{\includegraphics[width=0.8\linewidth]{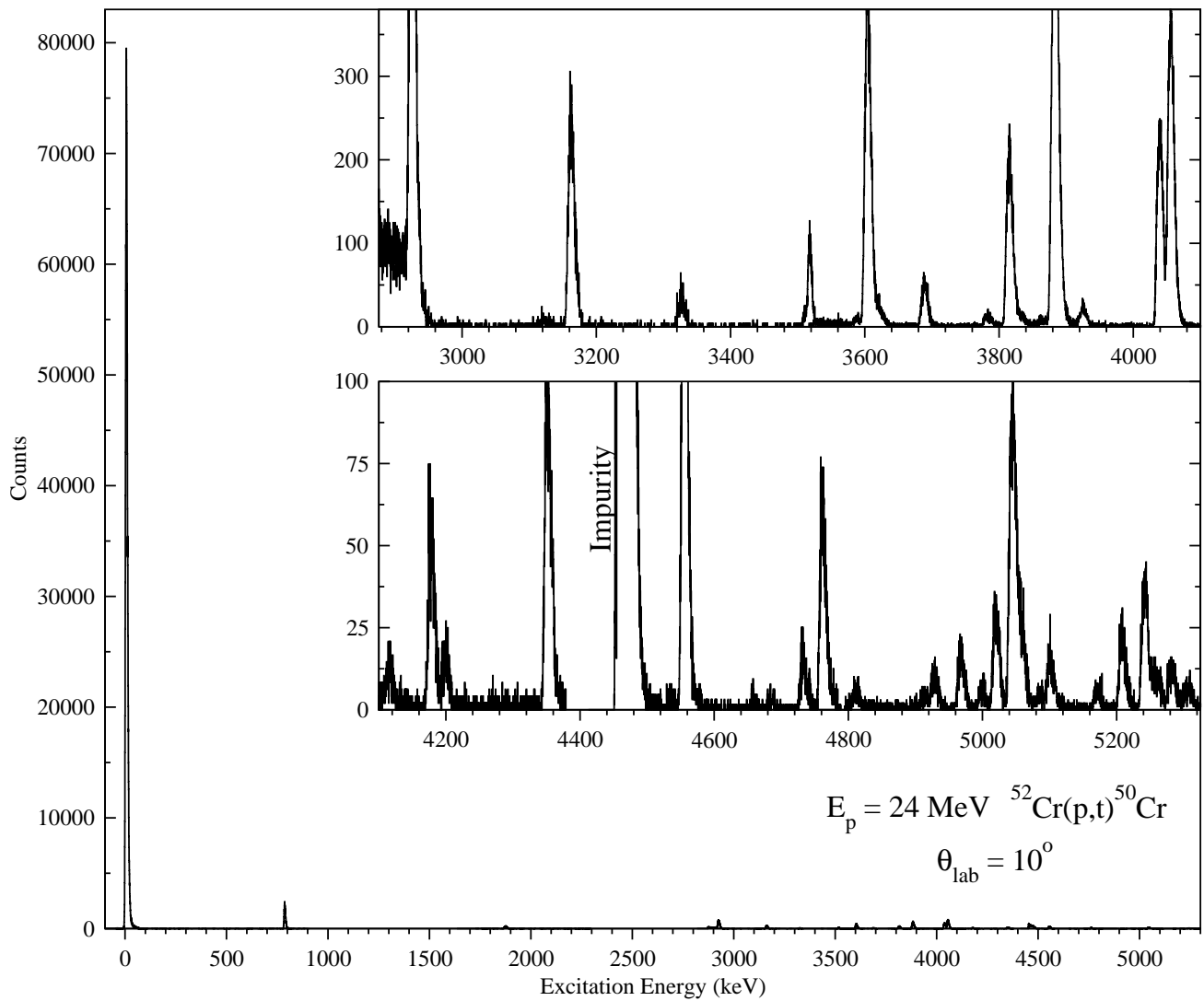}}
\caption{\label{full_ptspec50Cr}Spectrum of observed states in $^{50}$Cr at $\theta_{\rm{lab}}=10^{\circ}$ resulting from 24~MeV protons on $>99\%$ $^{52}$Cr.  Experimental limitations required states from 2.8 to 3.4~MeV to be shown at $\theta_{\rm{lab}}=50^{\circ}$.  In order to appropriately show the individual level detail, two inset panels are expanded on the regions from 2800-4100~keV (top inset) and 4100-5600~keV (bottom inset).  The wide features at $\sim$2.9~MeV and 4.5~MeV are the result of (p,t) reactions on a light impurity within the target.}
\end{figure*}

\section{Experiments}
\begin{figure*}[t!]
\centering
\rotatebox{-90}{\includegraphics[width=0.8\linewidth]{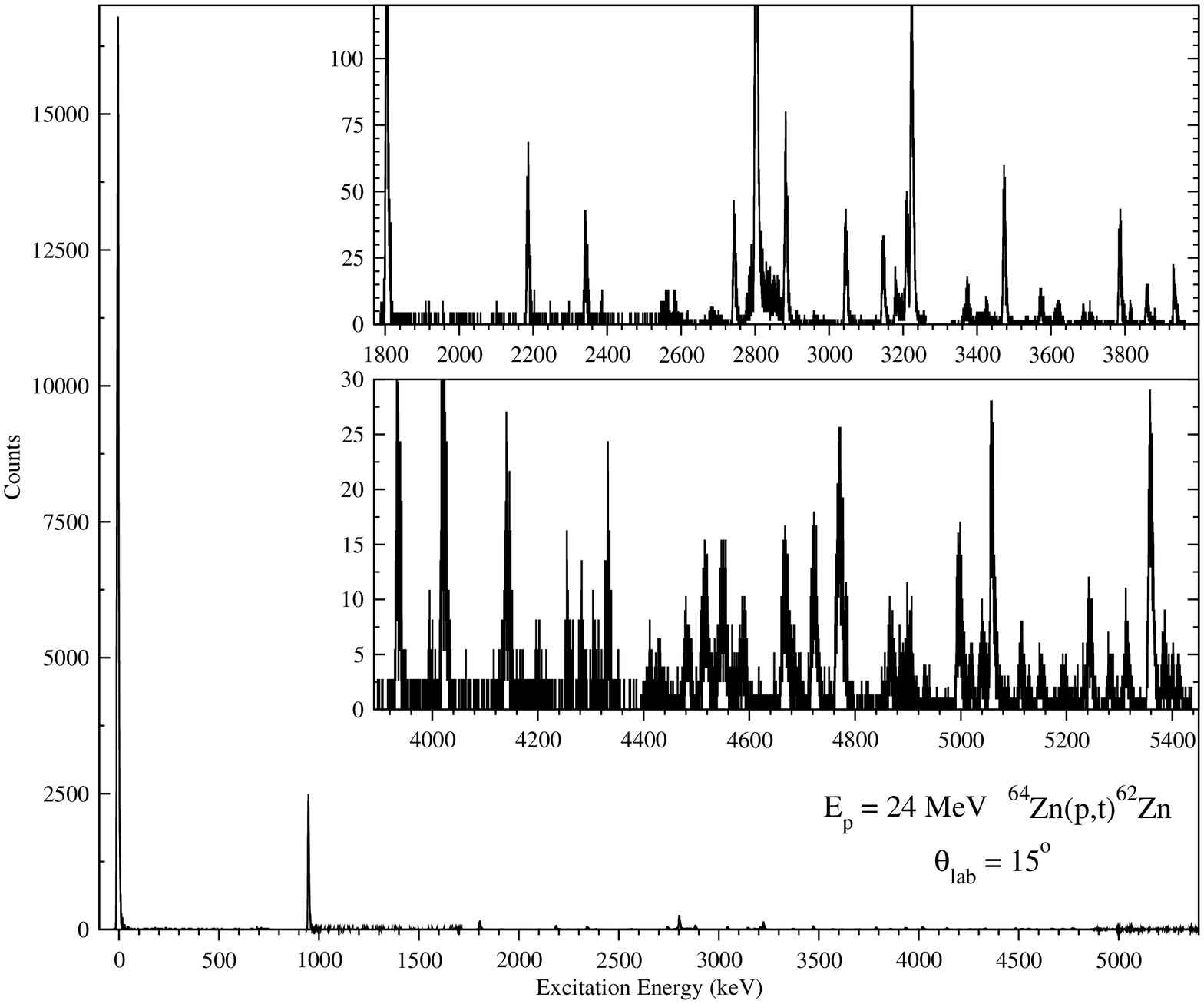}}
\caption{\label{full_ptspec}Spectrum of observed states in $^{62}$Zn at $\theta_{\rm{lab}}=15^{\circ}$ resulting from 24~MeV protons on $>99\%$ $^{64}$Zn.  Experimental limitations required states from 3.4 to 3.9~MeV to be shown at $\theta_{\rm{lab}}=20^{\circ}$, and states from 4.4 to 4.9~MeV to be shown at $\theta_{\rm{lab}}=25^{\circ}$.  In order to appropriately show the individual level detail, two inset panels are expanded on the regions from 1800-3900~keV (top inset), and 3900-5600~keV (bottom inset).  The wide feature at $\sim$2.8~MeV is the result of a (p,t) reaction on a light impurity within the target.}
\end{figure*}
The $^{52}$Cr(p,t) and $^{64}$Zn(p,t) experiments were both conducted at the Maier-Leibnitz-Laboratorium (MLL) of Ludwig-Maximilians-Universit\"at (LMU) and Technische Universit\"at M\"unchen (TUM) in Garching, Germany~\cite{Dol18}.  Beams of up to 1~$\mu$A of 24~MeV protons were incident on $>99\%$ isotopically pure, $\sim100~\mu$g/cm$^2$ targets with a 10~$\mu$g/cm$^2$ carbon backing.  The reaction products were momentum analyzed using a Q3D magnetic spectrograph, and the resulting particles were detected at the focal plane using a cathode-strip detector~\cite{Wir00} with a full-width at half-maximum (FWHM) energy resolution of roughly 10~keV.  For the $^{52}$Cr(p,t) experiment, outgoing tritons were observed at eight angles between $10^\circ$ and $50^\circ$, up to an excitation energy in $^{50}$Cr of 5.3~MeV using seven different momentum settings of the spectrograph.   For $^{64}$Zn(p,t), 9 angles between 10$^{\circ}$ and 60$^{\circ}$ were measured which required eight momentum settings of the spectrograph to cover excitation energies up to 5.4~MeV in $^{62}$Zn.  A $0^{\circ}$ Faraday cup inside the target chamber was used to determine the number of beam particles incident on the targets by integrating the total current.  This information, along with the data-acquisition system dead-time was read into the data stream using scalers.  Dead-time associated with the detector was also tracked, where all events gathered while the system was dead were binned in channel zero of each respective particle-energy spectrum.

Due to the large negative reaction $Q$ values for these (p,t) studies ($\sim-13$~MeV), no suitable calibration target exists to map the full energy response of the Q3D.  Therefore, to ensure the reported level energies were accurate, a careful examination of the triton energy dependence of the Q3D focal plane was conducted, and is reported previously in Refs.~\cite{Lea13a,Lea13b,Lea16}.  The calibrated energy spectra for observed states in $^{50}$Cr and $^{62}$Zn are displayed in Figs.~\ref{full_ptspec50Cr} and ~\ref{full_ptspec}, respectively.

Differential cross-sections were determined at each angle, for each momentum setting, using i) the integrated beam current at $0^{\circ}$, ii) the total solid angle of the magnetic spectrograph, iii) the respective target thicknesses, iv) system and detector dead-times, and v) the respective peak areas from the various energy spectra.  Using the complete set of measured cross-sections, angular distributions were constructed for all levels observed in both systems.  A 5\% systematic uncertainty was combined in quadrature with the other experimental and statistical uncertainties, to account for differences in the measured target thickness and other systematic effects, which was the limiting uncertainty in some cases.
\begin{figure*}[t!]
\centering
\rotatebox{-90}{\includegraphics[width=0.4\linewidth]{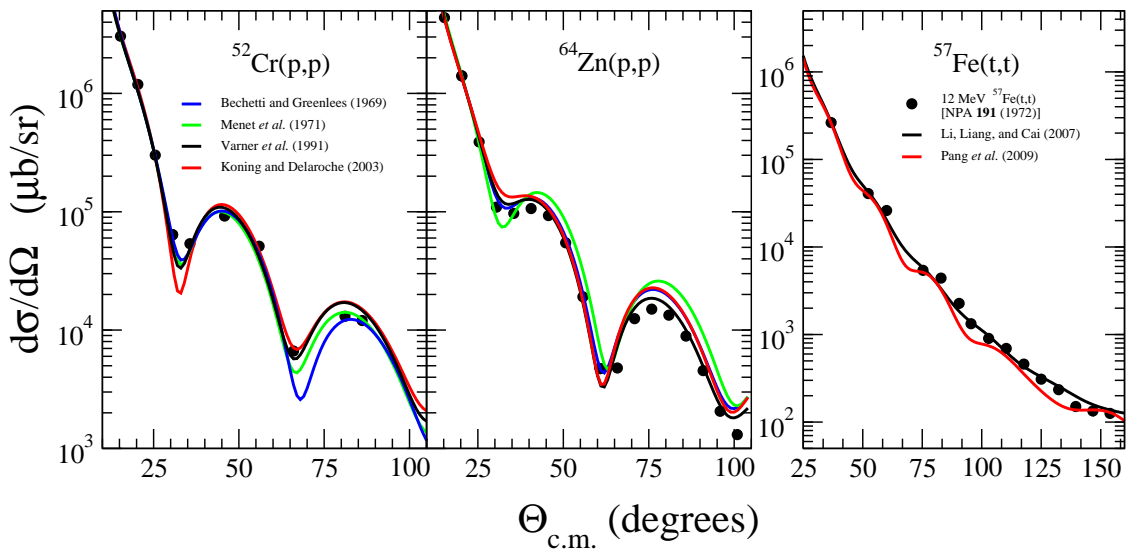}}
\caption{\label{pptt_fig}(Color online) Comparison of several global OMP sets with experimental data, for (left) 24~MeV proton elastic scattering from $^{64}$Zn, and (right) 12~MeV tritons off $^{57}$Fe(t,t)~\cite{McL72}.  All of the examined sets effectively reproduce the observed data, however the sets of Varner \textit{et al.}~\cite{Var91} and Li, Liang, and Cai~\cite{Li07} were chosen as the optimal sets, for protons and tritons, respectively.  The DWBA calculated curves performed using the optimal sets are shown in black.}
\end{figure*}

\section{DWBA Calculations}
The distorted-wave Born-approximation (DWBA) calculations presented here were performed using the finite-range, coupled-channel DWBA software package \textsc{fresco}~\cite{FRESCO}.  To obtain a better reproduction of the shape of the respective angular distributions for the purpose of level assignments, contributions from various shell configurations were accounted for in the \textsc{fresco} calculations of even parity states using shell-model two-nucleon amplitude (.tna) files~\cite{BroPC}.  These were particularly important for reproducing the $0^+$ states, as described in detail in Refs.~\cite{Lea13b,Lea16}.

The optical potential for the Born-approximation within the context of these calculations is defined as:
\begin{eqnarray}
V(r)&=&-V_{\mathrm{v}}f_{\mathrm{v}}(r)-iW_{\mathrm{v}}f_{\mathrm{v}}(r)+i4W_{\mathrm{s}}a_{\mathrm{s}}\frac{df_{\mathrm{s}}(r)}{dr}\nonumber \\
&&+\lambda_{\pi}^2\frac{V_{\mathrm{so}}+W_{\mathrm{so}}}{r}\frac{df_{\mathrm{so}}(r)}{dr}{\vec \sigma}\cdot{\vec \lambda}+V_C(r),\label{dwba_eqn}
\end{eqnarray}
\noindent where $V_C$ is the Coulomb potential defined as
\begin{eqnarray}
V_C(r)&=&\frac{ZZ'e^2}{r}~\mathrm{for}~r\geq R_{\mathrm{c}}\nonumber \\
&=&\left(\frac{ZZ'e^2}{2R_{\mathrm{c}}}\right)\left(3-\frac{r^2}{R^2_{\mathrm{c}}}\right)~\mathrm{for}~r\leq R_{\mathrm{c}},\label{dwba_eqn2}
\end{eqnarray}
\noindent
and where $R_{\mathrm{c}}=r_{\mathrm{c}}A^{1/3}$.  The volume term, $f_{\mathrm{v}}$, is of a Woods-Saxon form, while both the surface, $f_{\mathrm{s}}$, and spin-orbit, $f_{\mathrm{so}}$, terms are defined as the derivative of the Woods-Saxon potential.

The DWBA calculation results are heavily dependent on the incoming and outgoing channels of the reaction, which therefore requires appropriate optical-model parameters (OMPs) for the proton and triton at this mass and energy.  In an attempt to reduce the model-dependent systematic uncertainties as much as possible, the OMPs were not tuned to reproduce the transfer data, but rather were chosen to reproduce the respective proton and triton elastic scattering channel angular distributions (Fig.~\ref{pptt_fig}).  Globally determined OMP sets for both particles were used to even further reduce the possibility of finding a local minimum in the respective potentials, and the selection criteria for the these global OMP sets are outlined below.

\subsubsection{Proton}
A 24~MeV proton elastic scattering measurement on $^{64}$Zn was also performed in this work at 18 angles between $15^\circ$ and $100^\circ$.  Four global OMP sets were compared: Becchetti and Greenlees (1969)~\cite{Bec69}, J. J. H. Menet \textit{et al.} (1971)~\cite{Men71}, R.L. Varner \textit{et al.} (1991)~\cite{Var91}, and Koning and Delaroche (2003)~\cite{Kon03}.  As a result of this comparison, the OMPs from Ref.~\cite{Var91} were chosen as the optimal parameters for 24~MeV protons.

\subsubsection{\label{tritondetermination}Triton}
For the outgoing reaction channel, experimental triton elastic scattering data are not directly available.  Therefore, an examination of the published data closest to the appropriate energy and mass range was performed.  Using the Experimental Nuclear Reaction Data (EXFOR) database~\cite{EXFOR}, cross-sections for 12~MeV triton elastic scattering on $^{57}$Fe were found to be the closest available set of experimental data~\cite{McL72}.  Fewer triton OMP sets are available, however the chosen sets for comparison here are those of Li, Liang, and Cai (2007)~\cite{Li07}, and Pang {\it et al.} (2009)~\cite{Pan09}.  A comparison of the two global triton OMP sets to the experimental data was performed and is presented in Ref.~\cite{Lea13a}, and the optimal triton OMP set was determined to be that of Li, Liang, and Cai (2007)~\cite{Li07}.
\begin{figure}[t!]
\centering
\rotatebox{-90}{\includegraphics[width=0.6\linewidth]{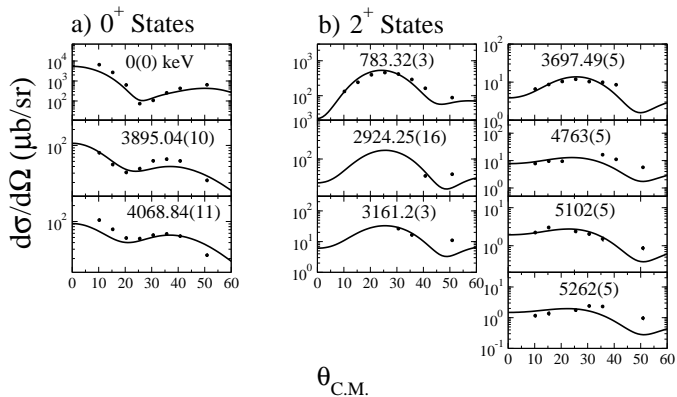}}
\caption{\label{0plus_2plus50Cr}Experimental angular distributions and FRESCO calculated curves for all observed a) $0^+$ states and b) $2^+$ states observed in $^{50}$Cr.  All energies listed are in keV.}
\end{figure}

\section{\label{pt_discussion}Experimental Assignments}
Observed $0^+$ states in both $^{50}$Cr and $^{62}$Zn from this experimental campaign are previously reported in Refs.~\cite{Lea16} and~\cite{Lea13b}, respectively, along with implications for the ``Towner-Hardy" theoretical shell-model isospin mixing corrections to superallowed $\beta^+$ decay~\cite{Tow08,Har15}.  Further detail and discussion on the observed states for $L\geq1$ two-neutron transfers is given in the sections below.  The full list of observed levels in $^{50}$Cr and $^{62}$Zn, including measured energies, spin-parity assignments, shell-model DWBA wave functions, differential cross-sections, and comparisons to previous observations are displayed in Tables~\ref{tab:50pt_info} and \ref{tab:62pt_info}, respectively.

\begin{figure}[t!]
\centering
\rotatebox{-90}{\includegraphics[width=0.7\linewidth]{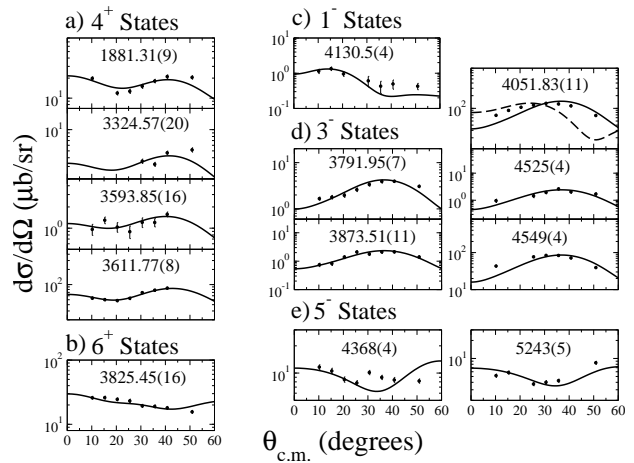}}
\caption{\label{other_states50Cr}Experimental angular distributions for all observed a) $4^+$, b) $6^+$, c) $1^-$, d) $3^-$, and e) $5^-$ states in $^{50}$Cr.  All energies listed are in keV.}
\end{figure}

\begin{figure}[t!]
\centering
\rotatebox{-90}{\includegraphics[width=0.7\linewidth]{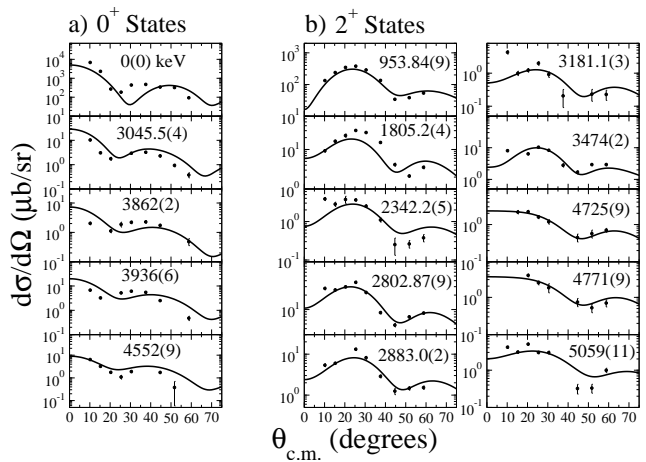}}
\caption{\label{0plus_2plus}Experimental angular distributions and FRESCO calculated curves for all observed a) $0^+$ states and b) $2^+$ states observed in $^{62}$Zn.  All energies listed are in keV.}
\end{figure}

The present results confirm many of the low-spin, natural-parity state assignments, as well as providing several new energy and spin assignments.  The measured angular distributions for states in $^{50}$Cr are shown in Figs.~\ref{0plus_2plus50Cr, other_states50Cr}, and those for states in $^{62}$Zn are shown in Figs.~\ref{0plus_2plus,other_states62Zn}, along with the \textsc{FRESCO} calculations. However, due to low population cross-sections and the loss of information at several angles from impurity peaks, many of the angular distributions are rather featureless, thus making firm assignments difficult.  Further, the extremely low cross sections for many of these levels (at the sub-$\mu$b level), prevents the exclusion of multi-step processes (Fig.~\ref{unassigned_states}).  The result is that, while natural-parity of the observed levels is favoured, it cannot be guaranteed.

\subsection{$^{50}$Cr}

{\bf 1$^-$ States} - Only one $1^-$ state was assigned in the work presented here, with an excitation energy in $^{50}$Cr of 4130.5(4)~keV.  This assignment removes the uncertainty in the quoted spin/parity listed in the evaluated data~\cite{Ele11} of $(1,2^+)$ based on $^{50}$Cr(p,p') measurements.  No other $1^-$ states under 6~MeV are reported in the literature.

{\bf 2$^+$ States} - Seven $2^+$ states were firmly assigned in this work.  The states at 783.32(3)~keV, 2924.25(16)~keV, 3161.2(3)~keV, and 4763(5)~keV confirm previous assignments in the evaluated data.  A previously observed state at 3697.9(6)~keV was given possible spin assignments of $1^+,2^+,3^+$ from $^{50}$Cr(p,p') measurements, but has been evaluated to an assignment of $1^+$.  The work presented here disagrees with that assignment, and presents this state as a $J^\pi=2^+$ with an excitation energy of 3697.49(5)~keV.  Further, two previously unobserved $2^+$ states are reported here with excitation energies of 5102(5)~keV and 5262(5)~keV.

{\bf 3$^-$ States} - The evaluated data of Ref.~\cite{Ele11} contains two definitively assigned $3^-$ states in $^{50}$Cr under 6~MeV, at energies of 4051.7(3)~keV and 4546.2(7)~keV.  The work presented here confirms both of those assignments with measured energies of 4051.83(9)~keV and 4549(4)~keV, respectively.  Three additional $3^-$ states are reported here, at energies of 3791.95(6)~keV, 3873.51(9)~keV, and 4525(4)~keV, which disagree with previous tentative assignments.

{\bf 4$^+$ States} - Four $4^+$ states were observed in this work at energies of 1881.31(9)~keV, 3324.57(20)~keV, 3593.85(16)~keV, and 3611.77(8)~keV.  Three of the states agree with the firm spin/parity assignment in the evaluated data, and the state at 3593.85(16)~keV removes the ambuguity of multiple assignments given in the evaluation.

{\bf 5$^-$ States} - One $5^-$ state is quoted in the evaluated data at $E_x=4367.0(4)$~keV, which is confirmed in this experiment.  Also presented here is an additional $5^-$ state at 5243(5)~keV, which is previously unobserved.

{\bf 6$^+$ States} - The evaluation of Ref.~\cite{Ele11} lists a tentatively assigned $6^+$ state at 3825.53(17)~keV, which is confirmed in this work with a measured energy of 3825.45(13)~keV.

\begin{longtable}{ccc|cc}
\caption{A complete list of all observed levels in $^{50}$Cr, including measured energies, spin-parity assignments, and differential cross-sections at $10^\circ$.  For comparison, the literature values~\cite{Ele11} for both level-energies and spin-parities are also included.  For states where $10^\circ$ data is unavailable, $50^\circ(\ddagger)$ cross-sections are reported.  The energies for states deonted with (*) were taken directly from Ref.~\cite{Ele11}, and were not included in the energy calibration.}
\label{tab:50pt_info} \\
\hline \hline
\multicolumn{1}{c}{$E_{exp.}$ (keV)} &
\multicolumn{1}{c}{$J^{\pi}_{exp.}$} &
\multicolumn{1}{c}{d$\sigma$/d$\Omega$ ($\mu$b/sr)} &
\multicolumn{1}{|c}{$E_{lit.}$ (keV)} &
\multicolumn{1}{c}{$J^{\pi}_{lit.}$} \\ \hline\hline
\endfirsthead
\multicolumn{5}{c}{{\tablename} \thetable{} -- Continued} \\
\hline \hline
\multicolumn{1}{c}{$E_{exp.}$ (keV)} &
\multicolumn{1}{c}{$J^{\pi}_{exp.}$} &
\multicolumn{1}{c}{d$\sigma$/d$\Omega$ ($\mu$b/sr)} &
\multicolumn{1}{|c}{$E_{lit.}$ (keV)} &
\multicolumn{1}{c}{$J^{\pi}_{lit.}$} \\ \hline\hline
\endhead
\hline
\endfoot
\hline \hline
\endlastfoot
0(0)* & $ 0^+$ & 6524(327) & 0(0)* & $ 0^+$ \\
783.32(3)* & $ 2^+$ & 133(7) & 783.32(3)* & $ 2^+$ \\
1881.31(9)* & $ 4^+$ & 19.8(12) & 1881.31(9)* & $ 4^+$  \\
2924.25(16)* & $ 2^+$ & $40(2)^{\ddagger}$ & 2924.25(16)* & $ 2^+$ \\
3161.2(3)* & $ 2^+$ & $11.0(7)^{\ddagger}$ & 3161.2(3)* & $ 2^+$ \\
3324.57(20) & $ 4^+$ & $5.2(4)^{\ddagger}$ & 3324.57(22) & $ 4^+$ \\
3593.85(16) & $ 4^+$ & 0.96(17) & 3594.57(22) & $2^+,3,4^+$ \\
3611.77(8) & $ 4^+$ & 54(3) & 3611.4(3) & $ 4^+$ \\
3628.7(3) & $-$ & 2.1(4) & 3629.6(7) & $1^+$ \\
3697.49(5) & $ 2^+$ & 6.6(5) & 3697.9(6) & $ 1^+$ \\
3791.95(6) & $ 3^-$ & 1.67(14) & 3792.0(4) & $(4^+,5^+)$ \\
3825.45(13) & $ 6^+$ & 25.9(14) & 3825.53(17) & $(6^+)$ \\
3845.5(2) & $-$ & 1.21(13) & 3844.3(3) & $2^+,3,4^+$ \\
3873.51(9) & $ 3^-$ & 0.75(10) & 3875.14(19) & $(4^+,5,6^+)$ \\
3895.04(8) & $ 0^+$ & 73(4) & 3894.6(12) & $ 4^+$ \\
3935.8(5) & $-$ & 2.32(20) & 3937.6(6) & $2^+,3,4^+$ \\
4051.83(9) & $ 3^-$ & 68(4) & 4051.7(3) & $3^-$ \\
4068.84(9) & $ 0^+$ & 106(6) & 4070.1(12) & $(2,3)$ \\
4130.5(4) & $ 1^-$ & 1.14(16) & 4129.5(4) & $(1,3)$ \\
4192(4) & $-$ & 7.1(5) & 4192.6(7) & 2$^+$ \\
4213(4) & $-$ & 2.0(2) & 4207(7) & $-$ \\
4286(4) & $-$ & 0.15(14) & 4282(7) & $-$ \\
4368(4) & $ 5^-$ & 11.6(7) & 4367.0(4) & $5^-$ \\
4525(4) & $ 3^-$ & 0.96(13) & 4524.4(9) & $(4^+)$ \\
4549(4) & $ 3^-$ & 43(2) & 4546.2(7) & $3^-$ \\
4655(5) & $-$ & 0.39(11) & 4654.3(3) & $(0^+,1,2,3^+)$ \\
4681(5) & $-$ & 0.34(11) & 4676(7) & $2^+$ \\
4733(5) & $-$ & 1.74(18) & $4728$ & $0^+$ \\
4763(5) & $ 2^+$ & 7.8(5) & $4772(7)$ & $2^+$ \\
4810(5) & $-$ & 0.69(13) & $-$ & $-$ \\
4913(5) & $-$ & 0.45(10) & $-$ & $-$ \\
4931(5) & $-$ & 1.29(15) & $-$ & $-$ \\
4970(5) & $-$ & 1.84(18) & $-$ & $-$ \\
5000(5) & $-$ & 0.83(9) & $-$ & $-$ \\
5021(5) & $-$ & 4.1(3) & $-$ & $-$ \\
5046(5) & $-$ & 14.2(8) & $-$ & $-$ \\
5060(5) & $-$ & 2.9(3) & $-$ & $-$ \\
5086(5) & $-$ & 0.71(8) & $-$ & $-$ \\
5102(5) & $ 2^+$ & 2.23(16) & $-$ & $-$ \\
5172(5) & $-$ & 0.73(8) & $-$ & $-$ \\
5210(5) & $-$ & 2.97(20) & $-$ & $-$ \\
5243(5) & $ 5^-$ & 5.2(3) & $-$ & $-$ \\
5262(5) & $ 2^+$ & 1.16(12) & $-$ & $-$ \\
5283(5) & $-$ & 1.73(13) & $-$ & $-$ \\
5308(5) & $-$ & 0.97(9) & $-$ & $-$ \\
\hline
\end{longtable}

\subsection{$^{62}$Zn}
\begin{figure}[t!]
\centering
\rotatebox{-90}{\includegraphics[width=0.7\linewidth]{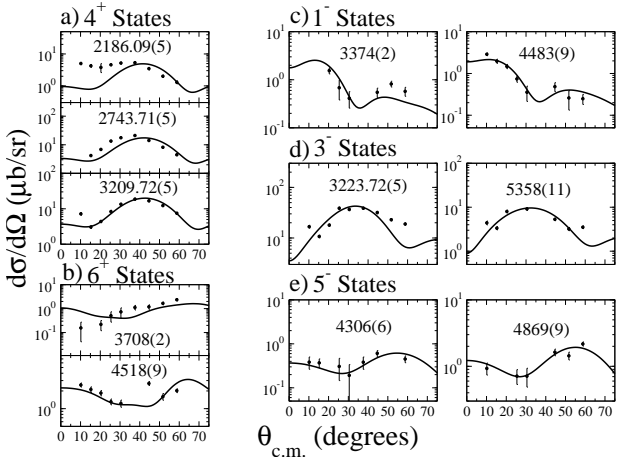}}
\caption{\label{other_states62Zn}Experimental angular distributions for all observed a) $4^+$, b) $6^+$, c) $1^-$, d) $3^-$, and e) $5^-$ states in $^{62}$Zn.  All energies listed are in keV.}
\end{figure}

{\bf 1$^-$ States} - Two $1^-$ states were populated in this work at 3374(2)~keV and 4483(9)~keV, respectively.  The lower of the two states was previously observed from the $^{62}$Ga $\beta$-decay work in Ref.~\cite{Fin08}, and given a tentative $(1^-,2^+)$ assignment.  This state is firmly assigned as a $1^-$ state from the agreement with DWBA curve in Fig.~\ref{other_states62Zn}.  The state at 4483~keV is also firmly assigned as a $1^-$ state, and is not previously reported in the literature.

{\bf 2$^+$ States} - In total, ten $2^+$ states were observed in this experiment.  Of these assigned states, most had been previously observed and a correlation of these states to those in Ref.~\cite{Nic12} was possible.  A tentatively assigned $(1^+)$ state at 3181~keV from the $^{62}$Ga superallowed $\beta$-decay work in Ref.~\cite{Fin08} was observed here, and is firmly assigned as a $2^+$ state from its characteristic angular distribution.  This assignment still agrees well with the data presented in Ref.~\cite{Fin08}, which is the only other previous observation of this state.  In addition, several states observed in this work above 5~MeV display $2^+$ characteristics, however large uncertainties prevent the firm assignment of these states.

{\bf 3$^-$ States} - Two previously observed states at 3223.72(5)~keV and 5358(11)~keV are firmly assigned as $3^-$ states from the work presented here.  The state at 3224~keV was tentatively assigned as $3^{\pm}$ by Ref.~\cite{Alb10} using $\gamma\gamma$ angular-correlation measurements, however a definite parity assignment could not be made.  Although this state is presented here as the $3^-_1$ state, recent heavy-ion reaction work in Ref.~\cite{Gel12} does not observe this level, but rather adopts a $3^-$ assignment for the 3209~keV state, despite the clear $4^+$ assignment of this state in Ref.~\cite{Alb10}.  The $3^-_2$ state at 5358~keV presented here agrees well with a 5370(20)~keV tentatively assigned $4^+$ state from previous (p,t) work, but is reassigned as $3^-$.

{\bf 4$^+$ States} - Three $4^+$ states were firmly assigned in this work, and correspond to previously known $4^+$ states listed in the evaluation in Ref.~\cite{Nic12}.  Recent high-spin work in Ref.~\cite{Gel12} assigns the 3210~keV state as the $3_1^-$, and does not agree with the $4^+_3$ assignment presented here and in the evaluated data~\cite{Nic12}.

{\bf 5$^-$ States} - Two $5^-$ states are presented here, neither of which have been observed previously.  The evaluation in Ref.~\cite{Nic12} lists one tentatively assigned $5^-$ state at 4043~keV, which is not observed in this work.

{\bf 6$^+$ States} - The well-known 3708~keV $6^+_1$ state from the ground-state-band was observed in this work, and confirms the previous 6$^+$ assignment.  The reported $6^+_2$ state at 4347.86(24)~keV was not observed in this experiment, however the $6^+_3$ state at 4515(20)~keV listed in the evaluated data was observed here with an energy of 4518(9)~keV.

\begin{longtable}[t!]{ccc|cc}
\caption{A complete list of all observed levels in $^{62}$Zn, including measured energies, spin-parity assignments, and differential cross-sections at $10^\circ$.  For comparison, the literature values~\cite{Nic12} for both level-energies and spin-parities are also included.  For states where $10^\circ$ data is unavailable, $15^\circ(\ddagger)$ or $20^\circ(\triangle)$ cross-sections are reported.  The energies for the ground-state and first excited state (*) were taken directly from Ref.~\cite{Nic12}, and were not included in the energy calibration.}
\label{tab:62pt_info} \\
\hline \hline
\multicolumn{1}{c}{$E_{exp.}$ (keV)} &
\multicolumn{1}{c}{$J^{\pi}_{exp.}$} &
\multicolumn{1}{c}{d$\sigma$/d$\Omega$ ($\mu$b/sr)} &
\multicolumn{1}{|c}{$E_{lit.}$ (keV)} &
\multicolumn{1}{c}{$J^{\pi}_{lit.}$} \\ \hline\hline
\endfirsthead
\multicolumn{5}{c}{{\tablename} \thetable{} -- Continued} \\
\hline \hline
\multicolumn{1}{c}{$E_{exp.}$ (keV)} &
\multicolumn{1}{c}{$J^{\pi}_{exp.}$} &
\multicolumn{1}{c}{d$\sigma$/d$\Omega$ ($\mu$b/sr)} &
\multicolumn{1}{|c}{$E_{lit.}$ (keV)} &
\multicolumn{1}{c}{$J^{\pi}_{lit.}$} \\ \hline\hline
\endhead
\hline
\endfoot
\hline \hline
\endlastfoot
0(0)* & $ 0^+$ & 6836(344) & 0(0)* & $0^+$ \\
953.84(9)* & $ 2^+$ & 133(7) & 953.84(9)* & $ 2^+$ \\
1805.2(4) & $ 2^+$ & 9.3(7) & 1804.67(11) & $ 2^+$ \\
2186.09(5) & $ 4^+$ & 5.0(4) & 2186.06(13) & $ 4^+$ \\
2342.2(5) & $0^+,2^+$ & 3.9(4) & 2341.95(23) & $0^+$ \\
2384.3(3) & $-$ & 0.45(11) & 2384.50(15) & $3^+$ \\
2743.71(5) & $ 4^+$ & 4.2(3)$^\ddagger$ & 2743.60(15) & $4^+$ \\
2802.87(9) & $ 2^+$ & 27.6(17) & 2803.14(17) & $2^+$ \\
2883.0(3) & $ 2^+$ & 5.4(4) & 2884.05(25) & $2^+$ \\
3045.5(4) & $ 0^+$ & 10.3(9) & 3042.9(8) & $0^+$ \\
3146.21(8) & $-$ & 2.5(2)$^\ddagger$ & $3160(10)$ & $(2^+)$ \\
3181.1(3) & $ 2^+$ & 4.3(5) & 3181.2(4) & $(1^+)$ \\
3209.72(5) & $ 4^+$ & 7.2(6) & 3209.86(21) & $4^+$ \\
3223.72(5) & $ 3^-$ & 16.7(11) & 3223.5(4) & $3^-$ \\
3374(2) & $ 1^-$ & 1.5(2)$^\triangle$ & 3374.2(3) & $1^-$ \\
3406(2) & $-$ & 0.22(17)$^\triangle$ & $-$ & $-$ \\
3443(2) & $-$ & 0.48(18)$^\triangle$ & $-$ & $-$ \\
3474(2) & $ 2^+$ & 8.1(6) & 3470(10) & $2^+$ \\
3571(2) & $-$ & 0.70(15) & $-$ & $-$ \\
3583(2) & $-$ & 0.64(15) & $3590(10)$ & $(2^+)$ \\
3621(2) & $-$ & 0.94(16) & $-$ & $-$ \\
3689(2) & $-$ & 0.43(13) & $-$ & $-$ \\
3708(2) & $ 6^+$ & 0.16(12) & 3707.60(24) & $6^+$ \\
3788(2) & $-$ & 7.5(5) & $-$ & $-$ \\
3817(2) & $-$ & 0.39(13) & $-$ & $-$ \\
3862(2) & $ 0^+$ & 2.0(2) & $3870(10)$ & $1^-$ \\
3884(2) & $-$ & 0.12(13) & $-$ & $-$ \\
3936(6) & $ 0^+$ & 6.8(6) & 4008.4(7) & $0^+$ \\
3994(6) & $-$ & 0.32(12) & $-$ & $-$ \\
4021(6) & $-$ & 5.8(5) & $4021.6(5)$ & $(1^+)$ \\
4141(6) & $-$ & 3.5(3) & $-$ & $-$ \\
4200(6) & $-$ & 0.21(11) & $-$ & $-$ \\
4218(6) & $-$ & 0.16(10) & $4217.6(8)$ & $3^-$ \\
4257(6) & $-$ & 0.83(16) & $-$ & $-$ \\
4282(6) & $-$ & 0.69(15) & $-$ & $-$ \\
4306(6) & $ 5^-$ & 0.38(12) & $-$ & $-$ \\
4331(6) & $-$ & 1.8(3) & $4330(20)$ & $(2^+)$ \\
4413(9) & $-$ & 0.41(15) & $-$ & $-$ \\
4432(9) & $-$ & 0.31(15) & $-$ & $-$ \\
4483(9) & $ 1^-$ & 2.9(2) & $4448.0(3)$ & $(1^+)$ \\
4518(9) & $ 6^+$ & 2.5(2) & 4515(20) & $6^+$ \\
4544(9) & $-$ & 0.77(13) & $-$ & $-$ \\
4552(9) & $ 0^+$ & 6.4(4) & 4620(20) & $0^+$ \\
4576(9) & $-$ & 0.30(11) & $-$ & $-$ \\
4590(9) & $-$ & 0.75(13) & $-$ & $-$ \\
4670(9) & $-$ & 1.65(17) & $4680(10)$ & $4^+$ \\
4688(9) & $-$ & 0.36(12) & $-$ & $-$ \\
4725(9) & $ 2^+$ & 2.2(2)$^\ddagger$ & $-$ & $-$ \\
4771(9) & $ 2^+$ & 3.9(3)$^\triangle$ & $4810(30)$ & $(2^+,3^-)$ \\
4778(9) & $-$ & 0.40(20)$^\triangle$ & $-$ & $-$ \\
4869(9) & $ 5^-$ & 0.93(19) & $4860(30)$ & $(3^-,4^+)$ \\
4894(9) & $-$ & 0.36(19) & $4895.3(4)$ & $(1^+)$ \\
4905(11) & $-$ & 0.8(2) & $4904.7(3)$ & $(7^-)$ \\
4933(11) & $-$ & 0.2(2) & $-$ & $-$ \\
4997(11) & $-$ & 2.1(2) & $-$ & $-$ \\
5018(11) & $-$ & 0.46(17) & $-$ & $-$ \\
5040(11) & $-$ & 0.80(18) & $-$ & $-$ \\
5059(11) & $ 2^+$ & 4.1(3) & 5050(30) & $2^+$ \\
5115(11) & $-$ & 0.49(17) & $-$ & $-$ \\
5151(11) & $-$ & 0.40(16) & $-$ & $-$ \\
5194(11) & $-$ & 0.64(19) & $-$ & $-$ \\
5243(11) & $-$ & 1.9(2) & $5240(20)$ & $0^+$ \\
5282(11) & $-$ & 0.76(20) & $-$ & $-$ \\
5313(11) & $-$ & 0.9(2) & $-$ & $-$ \\
5358(11) & $ 3^-$ & 4.4(4) & 5370(20) & $(4^+)$ \\
5387(11) & $-$ & 1.2(3) & $-$ & $-$ \\
5410(11) & $-$ & 0.2(3) & $-$ & $-$ \\
\hline
\end{longtable}

\subsection{Unassigned States}
\begin{figure}[t!]
\centering
\rotatebox{-90}{\includegraphics[width=\linewidth]{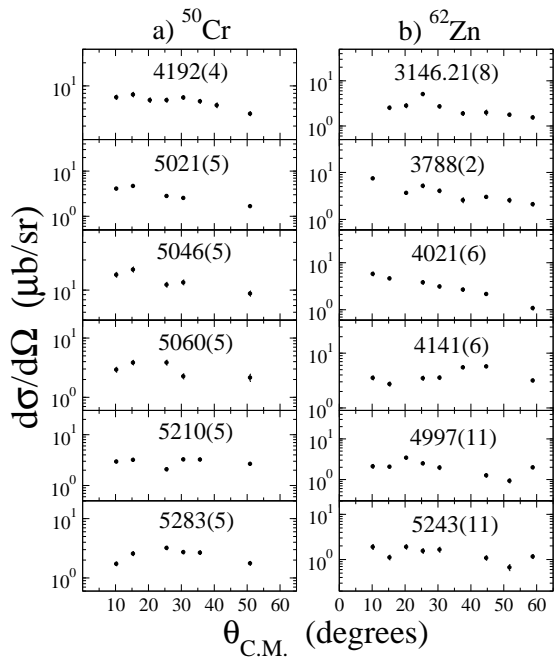}}
\caption{\label{unassigned_states}Experimental angular distributions for the six strongest unassigned states in a) $^{50}$Cr and b) $^{62}$Zn.  The relatively flat angular distributions likely result from a multi-step transfer of the two neutrons which removes the detail of the final-state nuclear structure.  All energies listed are in keV.}
\end{figure}

Due to the low population cross-sections, large reaction $Q$ values, and multi-step nature of the state populations, spin-parity assignments of many observed levels was not possible.  Of the states with reaction cross-sections above $1~\mu$b/sr which were not assigned, the six strongest are shown in Fig.~\ref{unassigned_states} to demonstrate the lack of structure in the angular distributions.  The data displayed in Fig.~\ref{unassigned_states} are typical of those for states which were populated more weakly as well.

\section{\label{discussion}Discussion}
\subsection{{\it ab-initio} Calculations using VS-IMSRG}
The {\it ab-initio} calculations presented here are based on $NN$ and $3N$ forces derived from~$\chi$-EFT~\cite{Epe09,Mach11PR}. In particular, the valence-space formulation~\cite{Tsuk12SM,Bogn14SM,Stro16TNO} of the in-medium similarity renormalization group (VS-IMSRG)~\cite{Herg16PR} is applied, which allows for tests of the nuclear forces in fully open shell nuclei. In the VS-IMSRG approach, an approximate unitary transformation~\cite{Morr15Magnus} is constructed to first decouple the $^{40}$Ca core, as well as an $pf$ valence-space Hamiltonian (comprised of the proton and neutron $f_{7/2}$, $f_{5/2}$, $p_{3/2}$, $p_{1/2}$ single-particle orbitals), diagonalized using the NuShellX@MSU shell-model code~\cite{Brow14NuShellX}. In the current IMSRG(2) level of approximation, throughout the calculation, all operators are truncated at the two-body level.  Bulk effects of $3N$ forces between valence nucleons are accounted for using the ensemble normal-ordering procedure of Ref.~\cite{Stro17ENO}. This procedure subsequently produces a distinct Hamiltonian for each nucleus under consideration. With four and ten protons outside the nearest closed shell, spectroscopy of the chromium and zinc isotopes, respectively, are currently beyond the reach of other large-space {\it ab-initio} methods.  The work presented here uses a particular set of $NN$ and $3N$ forces (the 1.8/2.0 (EM) interaction~\cite{Hebe11fits,Simo16unc,Simo17SatFinNuc}), which is fit in few-body systems but predicts ground state energies accurately to $^{100}$Sn~\cite{Morr18Sn}.

\subsection{Comparison to Experimental Data}
\begin{figure}[t!]
\centering
\includegraphics[width=\linewidth]{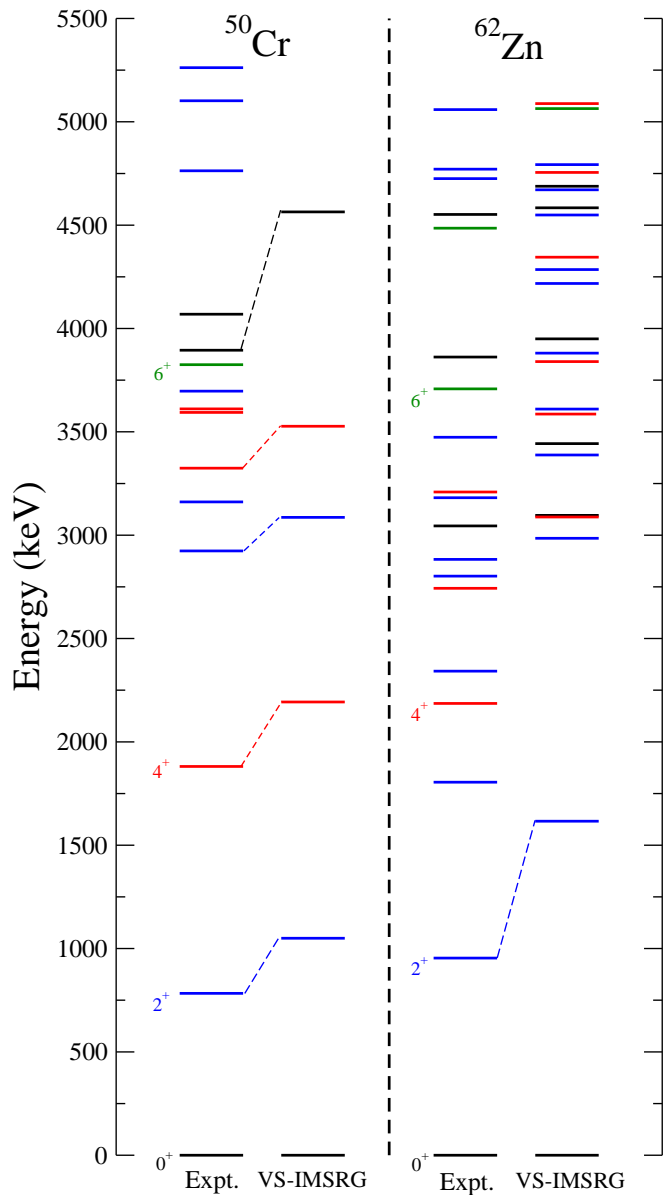}
\caption{\label{theory_compare}Comparison of the experimental $0^+$ (black), $2^+$ (blue), $4^+$ (red), and $6^+$ (green) states to the VS-IMSRG theoretical calculations for both $^{50}$Cr (left panel) and $^{62}$Zn (right panel).  Only the $2^+_1$ state in $^{62}$Zn has a connection line between experiment and theory to provide better clarity of the figure.}
\end{figure}

While ground- and excited-state energies of lighter isotopes in the $pf$ shell have been studied within the VS-IMSRG framework using the 1.8/2.0(EM) chiral interaction \cite{Leis18Ti,Moug18Cr,Izzo17Co,Step17Sc,Garn17M3}, this is the first attempt to study spectroscopy of mid-shell isotopes, which often exhibit highly collective features. Figure 9 shows the comparison between these {\it ab-initio} predictions and the new experimental data for even-pairty, even-$J$ states. In $^{50}$Cr, the low-lying states below 3~MeV are in reasonable agreement with experiment. While the lowest few states are higher in energy by 200-300~keV compared to experiment, the relative spacing between levels is quite similar. However above 3~MeV, the density of states is much lower than the observed experimental spectrum, which likely corresponds to states with dominant non-valence configurations. The situation in $^{62}$Zn is much more striking. Here both the first $2^+$ state and the grouping of higher-energy states are an MeV or more too high in energy. Nonetheless, the density of states in this case is more reasonable compared to experiment than in $^{50}$Cr.

It should be noted that because of the particular truncation scheme of many-body operators in the VS-IMSRG procedure, similar to other large-space {\it ab-initio} methods, we do not expect to fully capture the physics of highly collective nuclei~\cite{Hend18E2}. It appears that this is the case for these particular nuclei, with too spread spectra in both cases. We expect that improving the calculations to take into account in some way the effects of three-body operators, would result in more compressed spectra. 

\section{Conclusions}
This article presents detailed nuclear-structure measurements in the superallowed $\beta$-decay daughter nuclei $^{50}$Cr and $^{62}$Zn performed using 24~MeV (p,t) reactions.  Outgoing tritons were momentum-analyzed with the high-resolution Q3D magnetic spectrograph at the MLL in Garching, Germany, up to excitation energies of roughly 5.5~MeV.  Forty-five states in $^{50}$Cr and sixty-seven states in $^{62}$Zn were observed, including several previously unobserved or unassigned states.  The experimental spectroscopic results were also compared to {\it ab-initio} shell-model calculations using $\chi$-EFT with the VS-IMSRG method in the mid $pf$-shell for the first time, and show good agreement in the level orderings and density, albeit with slightly higher calculated energies for yrast states.  This characteristic of overbinding in the calculations is consistent with previous studies which demonstrate $\sim$ MeV-scale differences in the calculated atomic masses relative to experiment~\cite{Rei17,Leis18Ti,Moug18Cr,Izzo17Co,Step17Sc}.

As these theoretical techniques are further refined and the interactions are better understood, experimental benchmarking of these approaches in intermediate-mass systems are critical before they can be used for precision studies.  For ISB corrections to the measured $0^+\rightarrow0^+$ superallowed $\beta$ decay $ft$ values in particular, the next step is a reproduction of the measured $b$ and $c$ coefficients of the isobaric multiplet mass equation in the relevant $T=1$ isobaric triplets.

\section{Acknowledgements}
This work was supported in part by the U.S. Department of Energy, Office of Science under grant No. DE-SC0017649, the Natural Sciences and Engineering Research Council of Canada (NSERC), the Ontario Ministry of Economic Development and Innovation, the DFG Cluster of Excellence `Origin and Structure of the Universe'.  KGL would like to thank B. Alex Brown for providing the .tna files for the DWBA calculations, and Ian Thompson for his assistance with \textsc{fresco}.  The authors would also like to thank the MLL operation staff for providing a high-quality beam of 24~MeV protons during both experiments.

\bibliography{references}

\end{document}